\documentstyle[editedvolume]{crckapb} 

\def\gtsim{\lower.5ex\hbox{$\buildrel > \over\sim$}}
\def\ltsim{\lower.5ex\hbox{$\buildrel < \over\sim$}}

\begin{opening}
\title{NICMOS OBSERVATIONS OF LUMINOUS \&\protect\\
       ULTRALUMINOUS INFRARED GALAXIES}

\author{Aaron S. Evans}
\institute{California Institute of Technology 105-24\\
           Pasadena, CA 91125}

\end{opening}

\runningtitle{HST NICMOS Observations}

\begin{document}

\begin{abstract}

HST NICMOS observations of a sample of 24 luminous (LIGs: $L_{\rm IR}
[8-1000 \mu{\rm m}] = 10^{11.0-11.99} L_\odot$) and ultraluminous (ULIGs:
$L_{\rm IR} \gtsim 10^{12.0} L_\odot$) infrared galaxies are presented.
The observations provide, for the first time, high resolution HST imaging
of the imbedded 1.1 - 2.2 $\mu$m nuclear regions of these mergers.  All
but one of the ULIGs are observed to have at least one compact (50-200 pc)
nucleus, and more than half contain what appear to be blue star clusters.
The warm infrared galaxies (i.e., the transition sources) are observed to
have bright nuclei which account for most of the light of the galaxy.
This, combined with the tendency for the light of ULIGs to become more
centrally concentrated as a function of increasing wavelength, implies
that most of their energy is generated within a region 50-200 pc
across.

\end{abstract}

\vskip -0.4in
\section{Introduction}

There exists strong morphological and spectroscopic evidence that
ultraluminous infrared galaxies (ULIGs) are the by-products of the
merger/ interaction of molecular gas-rich galaxies, and that their extreme
luminosities result from circumnuclear starbursts and active galactic
nuclei (AGN). During the ultraluminous phase, the radiation from
the starburst/AGN, most of which is emitted at UV-to-optical wavelengths,
is absorbed by circumnuclear dust and re-emitted at infrared/submillimeter
wavelengths.

Of direct relevance to this conference is whether stars or AGN account for
the majority of the energy output of ULIGs. Based on the similarities in
the luminosity functions of ULIGs and QSOs, and the evidence for
morphological pecularities and large molecular gas masses in some QSOs,
Sanders et al. (1988a,b) proposed an evolutionary connection between ULIGs
and QSOs.  A population of ``warm'' ULIGs (i.e., $f_{25}/f_{60} > 0.2$,
similiar to the colors of Seyfert galaxies) were singled out as an
advanced stage of the ULIG phenomenon, occuring between the ``cool''
ultraluminous phase, when enhanced star formation is occuring and the AGN
is being built, and the optical QSO phase, when most of the gas and
dust have been consumed/blown out by star formation and the AGN.

\begin{figure}
\includegraphics{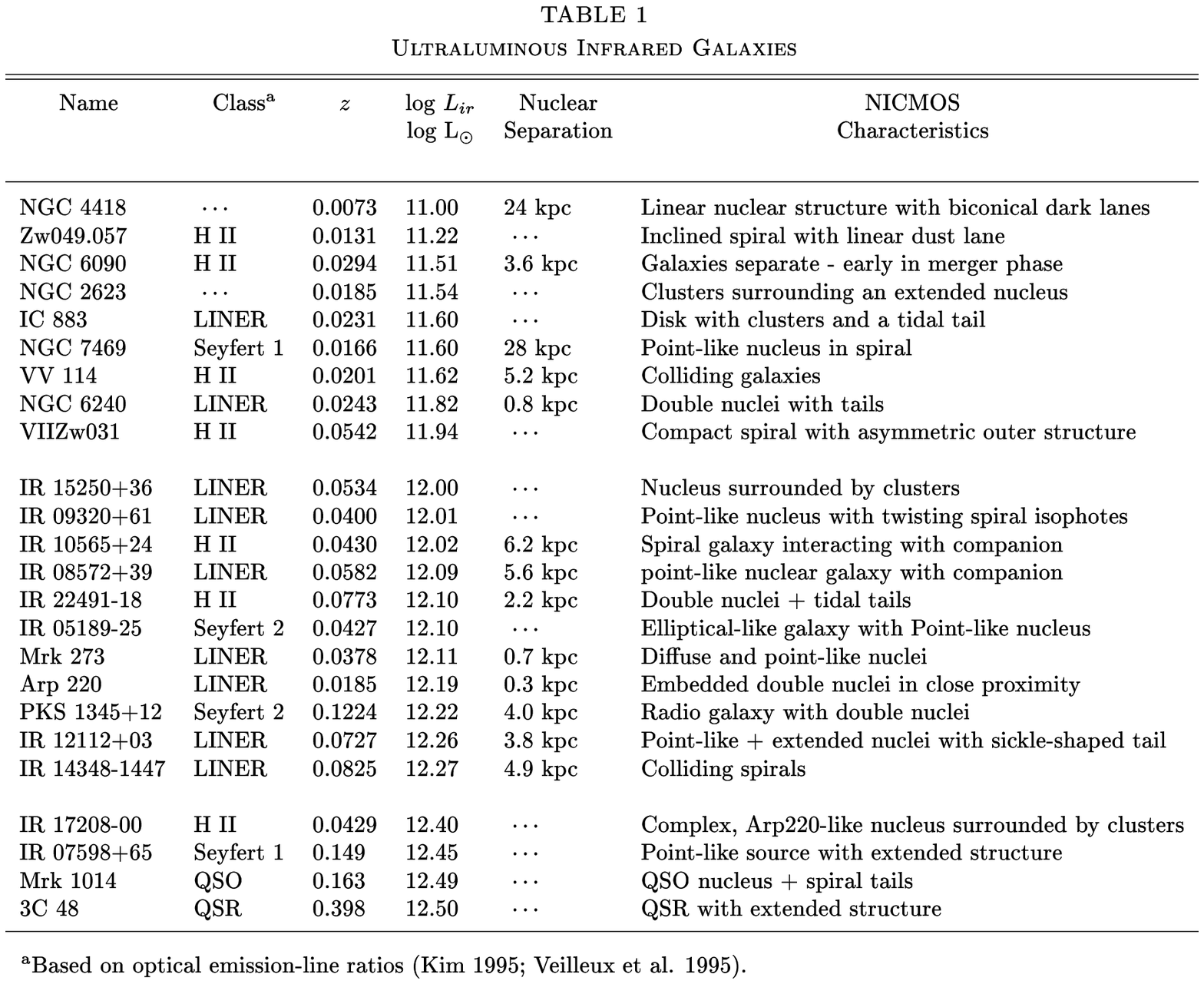}
\vspace{10cm}
\end{figure}

The high column densities intrinsic to ULIGs have made attempts to image
their nuclear regions at optical wavelengths difficult.  Observations with
ground-based arrays sensitive to radiation emitted at near-infrared
wavelengths (where the extinction is ten times lower than at optical
wavelengths) have improved our ability to observe the morphology and colors
of imbedded regions, but have been limited by atmospheric seeing (FWHM
$\approx 0.7^{\prime \prime}$: Murphy et al. 1996; Kim 1996).  The
availability of the Near-Infrared Camera and Multiobject Spectrometer
(NICMOS), recently installed on the Hubble Space Telescope (HST), thus
offers a unique opportunity to study ULIGs by combining near-infrared
technology with the high resolution and photometric accuracy possible with
a space-based instrument.

As part of NICMOS GTO program, a sample of 15 ULIGs (9 cool, 4 warm, 2
QSOs) were observed with Camera 2 of NICMOS at 1.1, 1.6, and 2.2
$\mu$m. For comparison, images of a sample of 9 LIGs were also obtained.
Camera 2 provides a $0.1^{\prime \prime}-0.2^{\prime \prime}$ resolution
(FWHM) at 1.1--2.2 $\mu$m and a 19$^{\prime \prime}$ field of view, which
correspond to a resolution of 25-200pc and projected field of view of 4-40
kpc for the redshift range of 0.01--0.15. The specific issues to be
addressed are: 1) Is there evidence of imbedded star clusters or
multiple/unresolved nuclei that may be responsible for the far infrared
emission?  2) Is there evidence for interactions/mergers in infrared
galaxies that were previously observed to show none?  3) Is there evidence
that a large fraction of the light from ULIGs emanates from increasingly
more compact regions as a function of increasingly wavelength, as might be
expected for an AGN?  4) How do the imbedded cluster colors
compare with the colors of the underlying galaxy? Below, each of these
issues is addressed. An $H_0 = 75$ km
s$^{-1}$ Mpc$^{-1}$ and $q_0 = 0.0$ are assumed throughout.

\section{Morphologies and Nuclear Luminosities}

Table 1 contains a general summary of the properties of the 24 galaxies
observed, and Figure 1 shows several examples of their near-infrared
morphologies. The statistics of the data are as follows: 22/24 galaxies
show morphological disturbances to varying degrees (note that the two
exceptions, IR 05189-25 and 07598+65, have large-scale, low
surface-brightness optical debris outside of the nuclear region: Surace et
al. 1998); 11/24 of the galaxies show evidence of spiral structure/arms;
11/24 have double nuclei with projected separations ranging from 0.3--7.0
kpc (two additional galaxies, NGC 4418 and 7469, have companion galaxies
$\approx$26 kpc away, i.e., out of the NICMOS field of view); 16/24 have
one or more blue star clusters.

\begin{figure}
\includegraphics{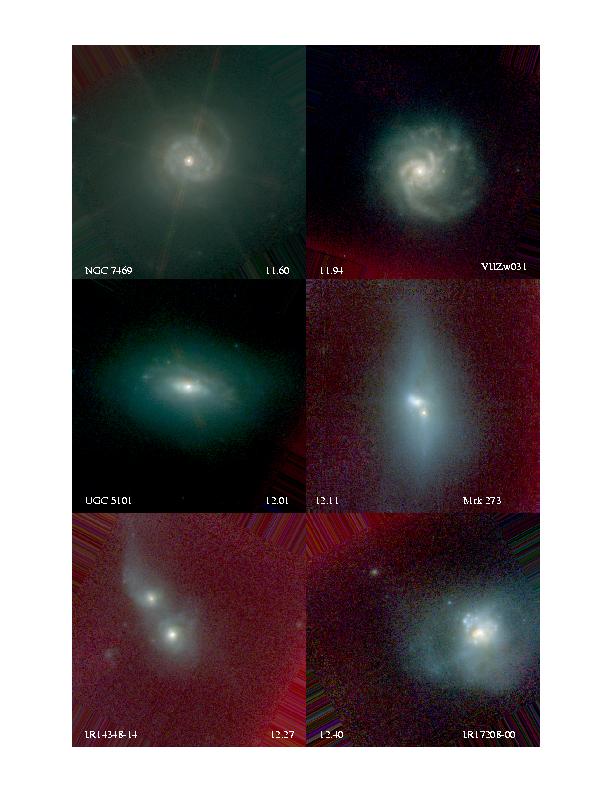}
\vspace{18cm}
\caption{NICMOS images of six of the LIGs/ULIGs. Blue=1.1 $\mu$m,
Green=1.6 $\mu$m, and Red=2.2 $\mu$m. The galaxy name and log$(L_{\rm IR})$
are provided in the bottom corners of each panel. The field of view is
$\approx 19^{\prime \prime} \times 19^{\prime \prime}$/panel.
North is up, east is to the left.} 
\end{figure}

In ground-based images of VIIZw031 (Figure 1) and IR 10565+24,
both have elliptical galaxy-like nuclear appearances (Djorgovski et
al. 1990; Murphy et al. 1996). In the NICMOS images, the central
0.5--1.0$^{\prime \prime}$ of the galaxies have spiral structure,
supporting the idea that the formation of a LIG or ULIG involves at least
one spiral galaxy. Similar nuclear spiral structure is also revealed
in HST images of NGC 7252 (Whitmore et al. 1993).

Putative nuclei with $L_{2.2\mu{\rm m}} \approx 2\times10^7 - 10^{11}$
L$_\odot$ have been identified in all galaxies.  Several of these
``nuclei'' have been detected in CO, and all but a few have strong
Pa$\alpha$ and radio emission. With the exception of IR 17208-00, all of
the ULIGs contain at least one compact (i.e., FWHM \ltsim 70-200 pc)
nucleus. The more compact cores tend to be the more luminous. In
particular, all of the warm ULIGs, whose nuclei at first glance look like
stars, have very luminous nuclei.

\section{Nuclear Concentrations}

Using 1.1$^{\prime \prime}$ and 11$^{\prime \prime}$-diameter apertures,
the percentage of the total galactic light contained within the nucleus of
the galaxies were measured. The smaller aperture size was selected to
contain the first airy disk of the unresolved 2.2 $\mu$m nuclei, thus
minimizing photometric errors due to the different resolutions of the 1.1,
1.6, and 2.2 $\mu$m images. The ULIGs with the lowest $f_{60}/f_{100}$
ratio (e.g. the cool ULIGs Arp 220, IR 12112+03) had the lowest value of
$f(1.1^{\prime \prime})/f(11^{\prime \prime}$) [$\approx$ 0.15-0.3 at 2.2
$\mu$m], whereas the warm ULIGs, which have high $f_{60}/f_{100}$ ratios,
have high values of $f(1.1^{\prime \prime})/f(11^{\prime \prime}$)
[$\approx$ 0.70].  The galaxy UGC 5101 (IR 09320+61: Figure 1), has an
intermediate value of $f(1.1^{\prime \prime})/f(11^{\prime \prime}$)
[=0.42], indicating that its 2.2 $\mu$m emission emanates from the nucleus
and underlying disk in nearly equal amounts.

Finally, $f(1.1^{\prime \prime})/f(11^{\prime \prime}$) increases as a
function of increasing wavelength in nearly all of the galaxies, possibly
indicating that their emission becomes more centrally concentrated at
longer wavelengths. This is due in part to the waning contribution of
light from the extended underlying stellar population at wavelengths
longward of 1.6 $\mu$m.

\section{Infrared Colors}

Figure 2 is a plot of $m_{1.1-1.6}$ vs. $m_{1.6-2.2}$ for all of the
galaxies in the sample, extracted using a $1.1^{\prime \prime}$-diameter
aperture.  Most of the LIGs and cool ULIGs are in a linear scatter with a
slope consistent with starlight reddened by a screen of cold dust,
however, a QSO embedded in warm ($> 1000$K) dust cannot be ruled out for
galaxies at the high end of the linear scatter of points (e.g., Arp 220,
IR 17208-00, IC 883, NGC 2623). Whether starlight or AGN, the nuclear
energy sources of these galaxies are heavily embedded.

\begin{figure}
\includegraphics{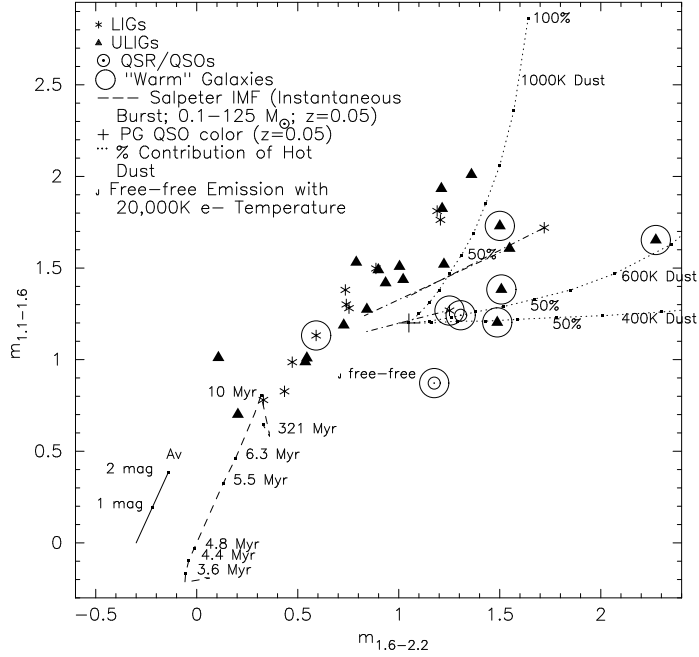}
\vspace{8.4cm}
\caption{Infrared Magnitude Color Diagram for LIGs/ULIGs measured in 1.1$^{\prime
\prime}$-diameter aperture. The dash-dotted line extending from UGC 5101
(triangle), VV 114 (*), and NGC 7469 (*) indicate how far the galaxies move on the
diagram when their colors are measured in 11$^{\prime \prime}$-diameter
apertures. Adapted from Scoville et al. (in prep).} 
\end{figure}

While all of the galaxies have $m_{1.1-1.6}$ colors between 0.7 and 2.0,
the warm ULIGs have high values of $m_{1.6-2.2}$ and lie on or to the
right of the QSO--1000K dust line. Thus, in addition to being extremely
compact and bright, the nuclei of the warm galaxies are also abnormally
red.  The similarities in the colors and nuclear compactness of warm ULIGs
and QSOs lends strong support to the idea that warm ULIGs possess
QSO nuclei.

Two LIGs and one cool ULIG also lie in the region of Figure 2 occupied by
the warm ULIGs.  One of the LIGs, NGC 7469, is a warm infrared galaxy
($f_{25}/f_{60}= 0.22$) with a bright Seyfert 1 nucleus embedded in a
spiral galaxy.  The cool ULIG, UGC 5101, contains a bright nucleus
embedded in a spiral galaxy.  In an 11$^{\prime \prime}$-diameter
aperture, both galaxies are indistinguishable from any of the other cool
ULIGs. Thus, UGC 5101 and NGC 7469 may be intermediate cases, possessing
warm, QSO-like infrared nuclei embedded in cool (and very luminous) 
infrared galaxies.

\section{Star Clusters}

A similar color analysis has been done of the clusters ($L_{2.2\mu{\rm m}}
\approx 10^4-10^7 L_{\odot}$) in a few of the galaxies. While no attempt
has been made to constrain the ages of the clusters, the cluster colors
appear to be bluer than the underlying galaxies, which are bluer than the
nuclei. If the assumption is made that the effects of extinction are
moderate or negligible outside of the nuclear region, the star clusters
are much younger than the underlying stellar population and are probably
by-products of the merger event.


\acknowledgements
It is a pleasure to thank my collaborators N. Scoville, N. Dinshaw, M.
Rieke, D. Hines, R. Thompson, J. Surace and the NICMOS GTO Team.
This research was supported NASA grant NAG 5-3042.

\end{document}